\documentclass[sigconf,natbib=true,anonymous=false]{acmart}
\usepackage[show]{chato-notes}
\usepackage{balance}
\usepackage{svg}
\AtBeginDocument{%
  \providecommand\BibTeX{{%
    \normalfont B\kern-0.5em{\scshape i\kern-0.25em b}\kern-0.8em\TeX}}}

\begin{document}


\title{Multimodal Neural Databases}

\author{Giovanni Trappolini}
\email{trappolini@diag.uniroma1.it}
\orcid{0000-0002-5515-634X}
\affiliation{%
  \institution{Sapienza University}
  \streetaddress{Via Ariosto, 25}
  \city{Rome}
  \state{Lazio}
  \country{Italy}
  \postcode{00185}
}

\author{Andrea Santilli}
\email{santilli@di.uniroma1.it}
\orcid{0000-0001-8483-2725}
\affiliation{%
  \institution{Sapienza University}
  \streetaddress{Via Ariosto, 25}
  \city{Rome}
  \state{Lazio}
  \country{Italy}
  \postcode{00185}
}

\author{Emanuele Rodolà}
\email{rodola@di.uniroma1.it}
\orcid{0000-0003-0091-7241}
\affiliation{%
  \institution{Sapienza University}
  \streetaddress{Via Ariosto, 25}
  \city{Rome}
  \state{Lazio}
  \country{Italy}
  \postcode{00185}
}

\author{Alon Halevy}
\email{ayh@meta.com}
\orcid{0000-0002-8717-7356}
\affiliation{%
  \institution{Meta AI}
  \streetaddress{One Hacker Way}
  \city{Menlo Park}
  \state{California}
  \country{USA}
  \postcode{94025}
}

\author{Fabrizio Silvestri}
\email{fsilvestri@diag.uniroma1.it}
\orcid{0000-0001-7669-9055}
\affiliation{%
  \institution{Sapienza University \& ISTI - CNR}
  \streetaddress{Via Ariosto, 25}
  \city{Rome}
  \state{Lazio}
  \country{Italy}
  \postcode{00185}
}


\renewcommand{\shortauthors}{Trappolini, et al.}



\begin{abstract}

The rise in loosely-structured data available through text, images, and other modalities has called for new ways of querying them.
Multimedia Information Retrieval has filled this gap and has witnessed exciting progress in recent years.
Tasks such as search and retrieval of extensive multimedia archives have undergone massive performance improvements, driven to a large extent by recent developments in multimodal deep learning.
However, methods in this field remain limited in the kinds of queries they support and, in particular, their inability to answer database-like queries.
For this reason, inspired by recent work on neural databases, we propose a new framework, which we name Multimodal Neural Databases (MMNDBs).
MMNDBs can answer complex database-like queries that involve reasoning over different input modalities, such as text and images, at scale.
In this paper, we present the first architecture able to fulfill this set of requirements and test it with several baselines, showing the limitations of currently available models.
The results show the potential of these new techniques to process unstructured data coming from different modalities, paving the way for future research in the area.
Code to replicate the experiments will be released at https://github.com/GiovanniTRA/MultimodalNeuralDatabases
  
\end{abstract}





\maketitle

\section{Introduction}

The amount and variety of data available have increased dramatically in recent years, and as more devices, such as smart glasses, become widespread, this trend is likely to accelerate. While current devices generate mostly text and image data, smart glasses will likely increase the amount of audio and video data individuals create. With the emergence of generative AI, we will likely see an explosion of valuable generated data. 
Multimedia Information Retrieval (MMIR) has always attracted the attention of scientists and practitioners in Information Retrieval. MMIR aims to address the challenges of processing queries on multimedia collections. Due to the enormous increase of data availability, MMIR has also seen a surge in its interest. The field has explored topics such as retrieval from large image archives, query by image, and retrieval based on face or fingerprint~\cite{bozzon2010multimedia}. 
However, this paper brings forward a novel and transformative idea: \emph{given the huge impact that the field of AI has having in all of the areas of technology, we argue that the MMIR field needs to explore systems that can handle more expressive database-like queries called multi-modal neural databases (MMNDBs)}.

We illustrate the potential of MMNDBs with an example. Consider the following query over an image archive: {\em how many images contain musical instruments?} Assume that the images in the collection are labeled with the objects that are identified in them (e.g., trumpet, avocado, person). 
Hence, an MMIR system is likely to be able to return images with trumpets, or other musical instruments. However, finding which objects are wind instruments (or a more detailed category) requires an additional reasoning step of a join with a database of instruments. Moreover, counting the number of images that satisfy our condition requires reasoning about the size of the answer set, an operation routinely done by database systems but not supported by MMIR systems. Examples can be more complicated, such as finding the most common musical instrument appearing in the photos or considering only photos taken in cities that hosted the Olympic games. As seen from the examples above, one of the critical needs of MMNDBs is the ability to reason about sets.

\begin{figure*}[t]
\centering
\includegraphics[width=0.7\textwidth]{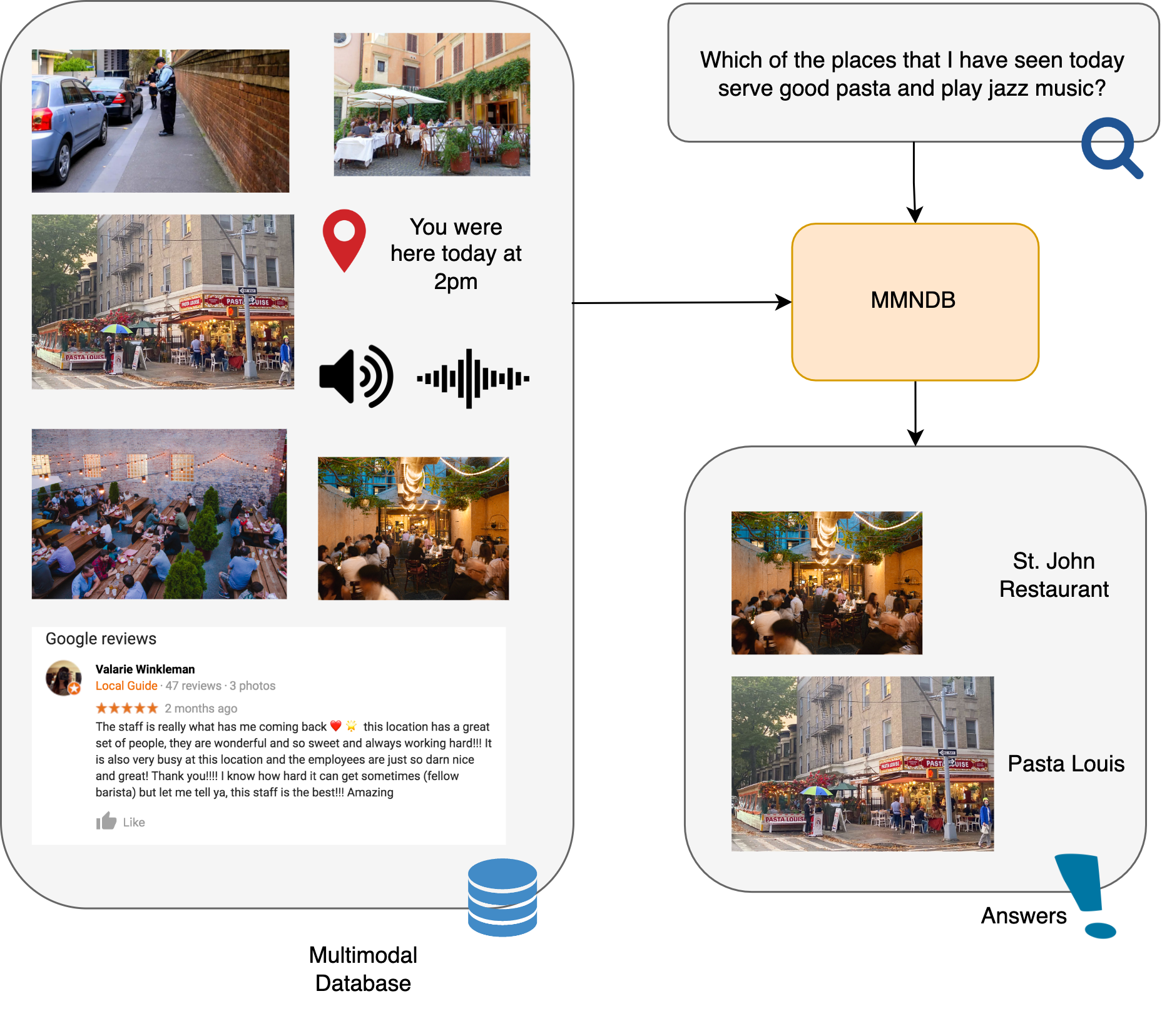}
\caption{A possible use case for MMNDBs. Imagine walking around the city with smart glasses and collecting information in a multimodal database. In the evening, you could be interested in knowing which are good places to eat that satisfy some criteria. MMNDBs could help make that decision by answering a database-like query posed in natural language (or voice!), combining multiple information sources and modalities.}
\label{fig:teaser_example}
\end{figure*}

In this perspective paper, we propose to study, design, and build MMNDBs by combining the capabilities of large multimodal models, multi-media information retrieval, and database query processing, as shown in Figure~\ref{fig:teaser_example}. We have been inspired by the work on neural databases~\cite{thorne2021natural,thorne2021database,sauchuk2022role} that have garnered interest in the NLP, database, and IR communities. However, we differentiate from that work as we position ourselves as an evolution of the field of MMIR by means of modern and, more recently proposed, multimodal AI technologies.

We develop a first principled prototype to show the proposed task's feasibility. We will later stress that this is only one of the possible architectures to solve MMNDBs and that future research will unveil new strategies. 
At a high level, we build our prototype on the retriever-reasoner-aggregator model. Given a query, the retriever returns a small subset of documents from the database that is relevant to the query. However, typically even that subset is too big to be provided as input to a single reader, which is essentially a transformer. Hence, the system runs multiple copies of the reasoner in parallel, each producing a partial result for the query. Finally, the aggregator component of the MMNDB will create the query result from the intermediate ones. For example, if the query counts the number of images that contain people, the intermediate results would be~1 or ~0, depending on whether the image contains a person. The aggregator will add up the~1s.

MMNDB systems will be designed to handle a wide range of multimedia data, including images, videos, audio, and text. 
The system will be able to process queries in natural language, allowing users to express their queries intuitively and easily. 
The system will also be able to extract features from multimedia data and use them to improve the performance of retrieval tasks.

\begin{figure*}[t]
\centering
\includegraphics[width=0.8\textwidth]{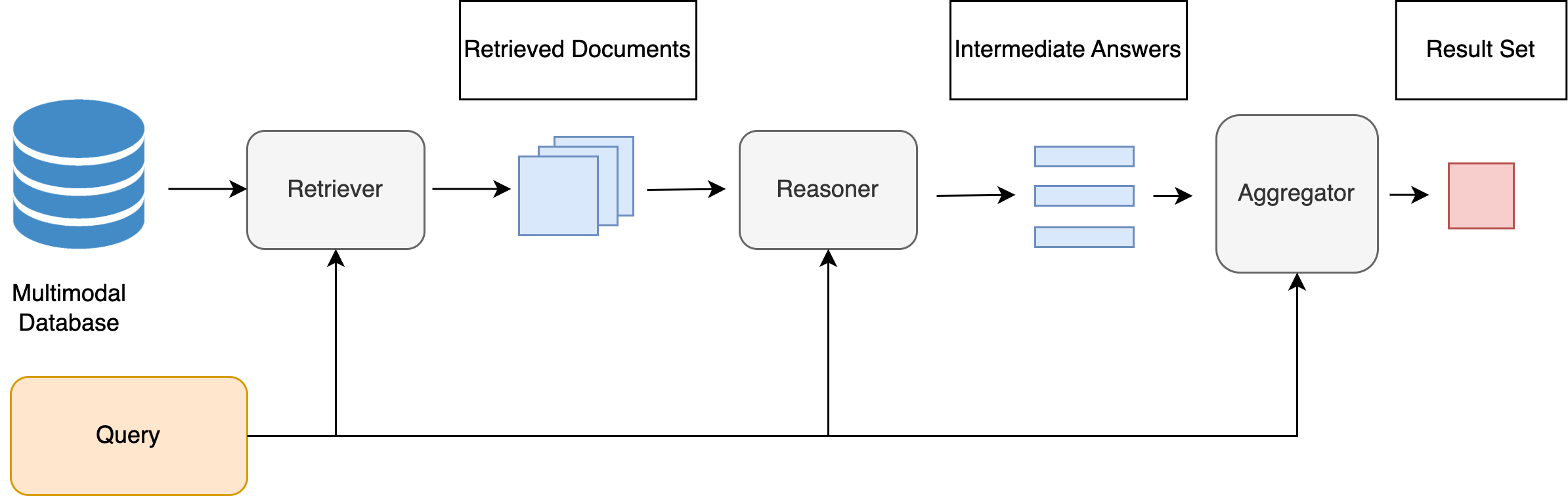}
\caption{Schema for our proposed MMNDB prototype. Given a query, documents are first filtered by a retriever module. A reasoner produces intermediate answers that are them processed by an aggregator to produce the final answer.}
\label{fig:architecture}
\end{figure*}

This paper describes a first step towards the realization of MMNDBs flexible enough to scaffold future models. We consider queries over collections of images and validate several aspects of our proposed architecture, as seen in Figure \ref{fig:architecture}. We perform a rich set of experiments that show the feasibility and potential of the proposed task across a subset of possible query types.
Finally, we discuss possible future research directions stemming from the anlysis brought forward in this paper and the introduction of Multimodal Neural Databases.

\section{Multimodal Neural Databases}

We refer to a corpus of \emph{documents} coming from different modalities as a multimodal database.
The definition of documents we provide here is intentionally very loose. In general, it could be any self-contained piece of data. 
Multimodal databases could include wildly different sources. For instance, it could contain information in natural language form, images, sounds, geo-tagging information, a timestamp, and many others.
Unlike a traditional database, a multimodal database is unstructured in the sense that it does not need to have a schema, or even less, it does not need to have any particular ordering but can be just the unordered and unstructured set of these documents.

Multimodal databases arise in several contexts. One existing context today is that of online social media, where users post content of different kinds (text, images, memes, videos, audio).
Here, each post is a document in the multimedia database, with the added peculiarity that the database would have to keep track of the graph of friendships between users. 
Another context that will arise in the near future when smart glasses are prevalent is the record of a user's day. Just by doing simple activities, like getting a coffee in a bar, the glasses will capture (adhering to whatever privacy conventions get adopted) sensory data, pictures (videos) of who is at the bar and what they are eating, audio of the background track playing, and photos of receipts for one's purchases.

Ideally, we would like to be able to query these rich, large, and unstructured collections of data the same way we query a database.
Going further, unlike a standard database, we would like to use natural language to perform queries instead of a rigid language like SQL.
Specifically, given a multimodal database $D$ and a query $q$, we would like to be able to perform the following types of query:
(a) \textbf{Set queries}; set queries are extractive queries that return a list of spans, such as entities, from the facts.
(b) \textbf{Boolean queries}; that return either True or False as an answer.
(c) \textbf{Join queries}; which require the combination of two or more documents to produce each answer.

We note that unlike traditional databases (or even neural databases), Multimodal databases can produce answers consisting of heterogeneous modalities. For instance, a set query can produce answers that include images, audio, and natural language (and their combination) seamlessly.

Designing a Multimodal Neural Database presents several substantial challenges.

First, it is crucial that the system is able to reason on the modalities given in input. For instance, if I were to look for images of cats and dogs fighting, I need to recognize both the presence of these animals \emph{and} that the interactions between the two is indeed that of fighting (a poster of Mike Tyson boxing in the background is not sufficient).
Similarly, if the query mentions someone whispering or yelling, the system must understand such subtleties in an audio frame.
Recently, deep learning techniques, particularly large deep learning models, have shown excellent reasoning capabilities \cite{goyal2017making}.
The tasks of Visual Question Answering and multi-hop question answering have reached near human results \cite{wang2018glue} for natural language processing, with promising candidates in the multimodal setting as well.

However, these models are usually extremely large, with billions of parameters, leading to the next challenge, namely scale.
Given a large collection of documents, it is infeasible to run such models on every query-document pair, or even on every document for that matter.

Open domain question answering systems (ODQA), developed for answering queries from natural language text,  provide a methodology for scaling to larger document collections. 
ODQA answers a query by first retrieving relevant documents from the document collection and feeding them as context to a transformer along with the query. 
However, transformers can only accept contexts of limited sizes (currently, 512 to 1024 tokens). Even though extending these sizes is a very active area of research, it will always likely be smaller than the size necessary to process the kinds of queries we are striving for. The number of documents that need to be processed for answering database queries can be arbitrarily big, as can the intermediate result sets. 
In contrast, ODQA systems usually consider queries whose answers are small and can be obtained by feeding just a few documents to the transformer.   Furthermore, a multimodal database is an {\em unordered} set of documents, so we cannot exploit any locality heuristic to retrieve the relevant documents.

Last but definitely not least, there is a challenge of bridging between the different modalities in a multimodal database. 
To answer queries over multimodal data, one has to process, reason, and combine information coming not only from different documents but also from documents expressed in different modalities.
The literature in natural language processing and computer vision has recently paved the way and achieved outstanding results in the field.
Multimodal models have followed, showing excellent results in the task of text-to-image, image-to-text, and text-to-music.
However, most multimodal models available today tackle either the text-visual or the text-audio tasks.
Combining multiple modalities, while not unexplored~\cite{baevski2022data2vec}, still needs additional research efforts to reach suitable levels to address the task at hand.
In particular, to suitably address the task of MMNDB, we would need a ``true" multimodal model, which can reason on any possible modality given as input.
For further discussion on this and other current limitations/future research directions, we refer to Section \ref{sec:futureR}.


\begin{center}
\begin{table*}
\caption{Comparison of different Retriever models under the ``Mixed" retrieval strategy. While CLIP's versions featuring resnets as a backbone have higher F1 and precision scores, ViT-based models achieve higher recall. We opt for the latter, as it allows the Reasoner module to receive as much relevant information as possible, ultimately reducing the final pipeline error.}
\label{table:retriever}
\begin{tabular}{@{}lcccccc@{}}
\toprule
\textbf{Model} & \multicolumn{1}{c}{\textbf{$\mu$F1}} & \multicolumn{1}{c}{\textbf{$\mu$Recall}} & \multicolumn{1}{c}{\textbf{$\mu$Precision}}  & \multicolumn{1}{c}{\textbf{F1}} & \multicolumn{1}{c}{\textbf{Recall}} & \multicolumn{1}{c}{\textbf{Precision}}  \\ \midrule
RN50 & 0.315 $\pm$ 0.002 & 0.819 $\pm$ 0.003 & 0.195 $\pm$ 0.002 & 0.320 $\pm$ 0.018 & 0.731 $\pm$ 0.035 & 0.302 $\pm$ 0.026 \\
RN50x4 & 0.424 $\pm$ 0.002 & 0.794 $\pm$ 0.003 & 0.290 $\pm$ 0.002 & 0.447 $\pm$ 0.022 & 0.717 $\pm$ 0.031 & 0.419 $\pm$ 0.027 \\
RN50x16 & $\mathbf{0.440 \pm 0.002}$ & 0.791 $\pm$ 0.003 & $\mathbf{0.305 \pm 0.002}$ & $\mathbf{0.478 \pm 0.023}$ & 0.710 $\pm$ 0.029 & $\mathbf{0.457 \pm 0.028}$ \\
RN50x64 & 0.331 $\pm$ 0.002 & 0.837 $\pm$ 0.003 & 0.206 $\pm$ 0.002 & 0.384 $\pm$ 0.019 & 0.759 $\pm$ 0.034 & 0.343 $\pm$ 0.025 \\
RN101 & 0.344 $\pm$ 0.002 & 0.873 $\pm$ 0.003 & 0.214 $\pm$ 0.002 & 0.388 $\pm$ 0.021 & 0.809 $\pm$ 0.028 & 0.317 $\pm$ 0.024 \\
ViT-B/32 & 0.378 $\pm$ 0.002 & 0.876 $\pm$ 0.003 & 0.241 $\pm$ 0.002 & 0.395 $\pm$ 0.018 & 0.813 $\pm$ 0.022 & 0.298 $\pm$ 0.019 \\
ViT-L/14 & 0.324 $\pm$ 0.002 & 0.931 $\pm$ 0.002 & 0.196 $\pm$ 0.001 & 0.329 $\pm$ 0.015 & 0.894 $\pm$ 0.018 & 0.219 $\pm$ 0.013 \\
ViT-L/14@336px & 0.337 $\pm$ 0.002 & $\mathbf{0.932 \pm 0.002}$ & 0.205 $\pm$ 0.002 & 0.347 $\pm$ 0.016 & $\mathbf{0.905 \pm 0.015}$ & 0.228 $\pm$	0.014 \\ \bottomrule

\end{tabular}
\end{table*}
\end{center}


\section{A first prototype for MMNDB} 

To demonstrate the feasibility MMNDBs, this section describes a first prototype of such a system, for a restricted case.
We consider databases in which all the documents are images, and queries, which are posed in natural language, can express COUNT, MAX, and IN. 
However, as we explain below, the architecture for our preliminary system can apply to broader settings as well. We hope that this architecture forms the basis for other approaches to MMNDBs. 

Our system takes an input query $q$ over a database $D$. It includes three components.
The first component is the retriever, which selects a subset of the documents in $D$ that are relevant to answer the query.
The second component is the reasoner, which processes, possibly in parallel, subsets of the retrieved documents. The reasoner provides a partial answer to the query.
The third component is an aggregation operator that synthesizes the answers provided by the reasoner to compute the final answer to the query.

The strength of our architecture is that it enables us to exploit recent advances in multimodal neural models when implementing the retriever and the reasoner. Specifically, these models are able to map multiple modalities into the same embedding space, and therefore reason about the contents of images and text together. For example, these models can identify objects in images and create textual captions that describe the main aspects of of the image.

Before we explain each of the components, we give an end-to-end overview of how a query is processed in our system.
Consider the query ``How many people are playing the guitar in a blue t-shirt on a beach". The reasoner considers a single image in $D$ and uses the latest neural methods to determine whether the image contains a person playing guitar on the beach. However, applying such powerful reasoning on each of the documents in $D$ is infeasible, so we use a retriever to filter to only a small subset of the images in $D$, $P(D,q)$. Multiple instances of the reasoner then are applied in parallel to the retrieved images in $P(D,q)$ to determine which image satisfies the query. In our example, if  an image satisfies the query, the reasoner returns~1 and otherwise~0. The aggregator then counts the number of~1's to return the final answer. We now describe each of the components. 

\noindent \textbf{Retriever}.
The goal of the retriever  is to return a subset $P(D,q)$ of documents from $D$ that are relevant to the query $q$.  The main requirement from the retriever is that it be scalable. While the reasoning we expect from the retriever is not at the same granularity as the reasoner, it should weed out the vast majority of irrelevant images. To retrieve documents that are relevant to the query, we encode both the query and the documents in the same latent embedding space. However, as noted earlier, it is important that the embedding of a document {\em not} be dependent on the query $q$, otherwise we would have to compute a new embedding for every document in $D$ for any given query. Hence, as we describe in Section~\ref{sec:models}, we consider several methods for embedding the documents in $D$ in a query independent way. 

\noindent \textbf{Reasoner}.
An instance of the reasoner $P(D,q)$ takes one of the documents in $D$ as input and returns an intermediate answer to our query $A_p$. In the example above, the reasoner returns either~1 or~0 depending on whether the image satisfies the conditions in the query. However, the intermediate result may be different. For a query such as "What is the maximum number of people in the images" the reasoner would return, for every image, the number of people in that image. As another example, for the query "what is the most common musical instrument seen in the database", the output of the reasoner would be the list and number of occurrences of each of the instruments it identified in the image.

 The crucial role of the reasoner is, precisely, to reason about the relationship between the image and the query. In our example, the reasoner needs to determine whether there is a person wearing a blue outfit, that the same person is the one playing the guitar, and that they are physically located on a beach. The reasoner leverages the recent advances in neural models that are able to perform such reasoning by embedding the image and text in the same latent space and generating textual captions of images. 
It is worth noticing, however, that these models compute a dynamic embedding of the query and of the image, that depends on both, i.e., $F(I | T) \neq F(I)$ and vice versa, where $I$ is the image, and $T$ is natural language (could be any two modalities).
This has profound computational implications. 
In fact, to be able to answer the query, one would need to process any possible $D, q$ pair.
Furthermore, since the query is known only at inference time, it is not possible to precompute the embeddings.
It is then clearly unfeasible to run the reasoner on the entire database.
For this reason, we introduce an additional module in our pipeline, namely the retriever.

\noindent \textbf{Aggregator}.
The Aggregator takes as input the query and the set of intermediate outputs from all the instances of the reasoners and produces the answer to the query. Conceptually, this component of the system is the simplest because the intermediate results need to be aggregated depending on the semantics of the query. In our example, the aggregator would count the number of images for which~1 was returned. For the query counting the total number of people, the aggregator would sum the intermediate results returned from the reasoners.

\section{Experiments}

This section describes the experiments we performed to validate the promise of our prototype. We begin by describing the experimental settings. 

\subsection{Experimental setup}

In this section, we outline the experimental setup utilized to verify the validity of our approach.

\paragraph{Dataset}

Our experiments use the MS-COCO dataset (Common Object in Context) \cite{lin2014microsoft}, which is the single most popular benchmark dataset in computer vision. 
We use the latest version made available by the authors.
The COCO dataset contains approximately 123K labeled images.
Each image is associated with 5 captions and is annotated with the objects that are identified in it. The objects are drawn from a collection of  
 1.5M object instances across 80 object categories.
The dataset is divided into train and eval subsets, containing   118K and 5K images, respectively. 
We use the train set to train/fine-tune our methods while we report our results on the eval set.

\paragraph{Queries}

We use the MS-COCO dataset to build our queries.
For the COUNT query type, we may ask a query of the type ``How many \{object\} are in the database?", where object can be any of the object category contained in the COCO dataset.
Similarly, for the MAX query type, we may be interested in the image of the dataset with most frequent annotation of a particular kind.
Finally, for the In query, we are interested in images whose annotations satisfy certain conditions.


\begin{center}
\begin{table*}
\small
\caption{Results performance on the query type count. The PerfectIR setting acts as an ideal upper bound, showing the full potential of the MMNDB framework. The Full Pipeline (Full), on the other hand, shows excellent accuracy and $\Delta$ error but a total error that, while being good, is not at the level of PerfectIR. We empirically show that this is not caused by noise introduced by the retriever module, as indicated by the excellent results achieved in the NoisyIR setting.
Instead, this is caused by damaging documents picked up by the retriever that trick the reasoner resulting in a large False Positives error and, ultimately, a large total error.}
\label{table:pipeline_count}
\begin{tabular}{@{}l|c|ccc|ccc|ccc@{}}
\toprule
& \multicolumn{4}{c|}{\textbf{Total Error $\downarrow$}} & \multicolumn{3}{c|}{\textbf{$\Delta$ Error $\downarrow$}} & \multicolumn{3}{c}{\textbf{Accuracy $\uparrow$}} \\ \midrule
\textbf{Stock} & \multicolumn{1}{c|}{Error} & \multicolumn{1}{c}{Error TP} & Error FP & Error FN & \multicolumn{1}{c}{$\Delta$ Error} &\multicolumn{1}{c}{$\Delta$ Error TP} & $\Delta$ Error FP & \multicolumn{1}{c}{Accuracy} & \multicolumn{1}{c}{Accuracy TP} & \multicolumn{1}{c}{Accuracy FP} \\ \midrule
Perfect IR & $\mathbf{0.46  \pm 0.07}$ & 0.46  $\pm$ 0.07 &N/A&N/A& 4.64  $\pm$ 1.94 & 4.64  $\pm$ 1.91 & N/A & 0.60  $\pm$ 0.02 & 0.60  $\pm$ 0.02 & N/A \\
Noisy IR & 0.77  $\pm$ 0.16 & 0.46  $\pm$ 0.07 & $\mathbf{0.31  \pm 0.15}$ & N/A & $\mathbf{2.66  \pm 1.04}$ & 4.64  $\pm$ 1.91 & $\mathbf{0.31  \pm 0.02}$ & $\mathbf{0.81  \pm 0.01}$ & 0.60  $\pm$ 0.02 & 0.92  $\pm$ 0.01 \\
Dmg. IR & 1.24  $\pm$ 0.32 & 0.46  $\pm$ 0.07 & 0.78  $\pm$ 0.32 & N/A & 4.22  $\pm$ 1.41 & 4.64  $\pm$ 1.91 & 2.31  $\pm$ 1.15 & 0.70  $\pm$ 0.02 & 0.60  $\pm$ 0.02 & 0.76  $\pm$ 0.02 \\ 
Full	& 1.27	$\pm$ 0.17	& $\mathbf{0.42	\pm 0.07}$	& 0.76	$\pm$ 0.13 &	0.09	$\pm$ 0.02	& 3.33	$\pm$ 1.16	& 4.83	$\pm$ 2.03 &	1.96 $\pm$	0.80	& 0.73 $\pm$	0.02 &	$\mathbf{0.61	\pm 0.02}$	& 0.75	$\pm$ 0.02 \\
\midrule
 \textbf{FTmodel} &  \multicolumn{10}{l}{}\\ \midrule
 Perfect IR & $\mathbf{0.14 \pm 0.01}$ & $0.14 \pm 0.01$ & N/A & N/A & $1.46 \pm 0.10$ & $1.46 \pm 0.10$ & N/A & $0.67 \pm 0.02$ & $0.67 \pm 0.02$ & N/A \\ 
Noisy IR & $0.22 \pm 0.01$ & $0.14 \pm 0.01$ & $0.08 \pm 0.01$ & N/A & $\mathbf{0.90 \pm 0.06}$ & $1.46 \pm 0.10$ & $0.43 \pm 0.05$ & $\mathbf{0.86 \pm 0.01}$ & $0.67 \pm 0.02$ & $0.93 \pm 0.01$ \\ 
Dmg. IR & $0.54 \pm 0.05$ & $0.14 \pm 0.01$ & $0.40 \pm 0.05$ &N/A& $1.25 \pm 0.08$ & $1.46 \pm 0.10$ & $1.04 \pm 0.09$ & $0.73 \pm 0.01$ & $0.67 \pm 0.02$ & $0.73 \pm 0.02$ \\ 
Full & $0.99 \pm 0.06$ & $\mathbf{0.11 \pm 0.01}$ & $0.79 \pm 0.06$ & $0.09 \pm 0.02$ & $1.10 \pm 0.07$ & $\mathbf{1.42 \pm 0.10}$ & $0.99 \pm 0.07$ & $0.72 \pm 0.01$ & $\mathbf{0.69 \pm 0.02}$ & $0.72 \pm 0.02$ \\
\bottomrule
\end{tabular}
\end{table*}
\end{center}

\paragraph{Models}
\label{sec:models}

We now describe the neural models we used throughout our experiment.

For the Reasoner, we employ OFA \cite{wang2022unifying}.
OFA is a deep learning model trained on a wide variety of multimodal (text and image) tasks, ranging from image captioning to image generation, showing great results on unseen tasks as well.
OFA is open-source (code and weights) and is currently one of the best-performing multimodal models.  
We test four different versions of OFA, namely medium, base, large, and huge, with the largest featuring close to 1B parameters.
OFA is a transformer-based model that builds a joint representation of the input, namely text and visual, that is used to generate a textual response.
We stress again the fact that, given adequate computational resources, this module of the pipeline is highly parallelizable, hence capable of producing intermediate answers in the span of a few seconds.

For the Retriever, we employed the CLIP model \cite{radford2021learning}.
These models are trained in an unsupervised, contrastive manner by matching captions and images.
They take either text or images and align them in a shared latent space that can be used for later inferences and to measure their distance, with similar image-caption pairs being close together.
We test on 8 different versions of CLIP, namely RN50, RN101, RN50x4, RN50x16, RN50x64, ViT-B/32, ViT-L/14, ViT-L/14@366px.
CLIP's salient feature is that the created embeddings are static, meaning they do not depend on the query.  
This allows us to pre-compute the embeddings for all images beforehand, meaning that only the embedding for the query has to be computed at inference time.
Once the embeddings are computed, a strategy is needed to select which documents are considered relevant (and passed to the reasoner) and which ones are not.
To do this, we craft three strategies:

\noindent (i) \textit{TopK}: in this case, we compute the dot product between the embeddings of the documents and the query, we sort them, and we select the TopK documents.

\noindent (ii) \textit{Threshold}: we compute the cosine similarities between the embeddings of the text and the images, and we return all the documents for which the cosine similarity is greater than a certain threshold $\tau$ that depends on the particular CLIP model we are using, lying in a range between $0.15$ and $0.4$.

\noindent (iii) \textit{Neural Selector}: here, we train a small neural network that, given the $q$ and $D$ embeddings, returns a binary outcome that indicates whether the document is relevant for the query or not and whether it should be returned.

\noindent The actual number of parameters depends on the CLIP version employed, but it is always in the order of thousands.
It is worth noticing that, while it is still much more scalable with respect to the large 1B parameters models the reasoner employs, this strategy requires a ``dynamic" processing; namely, the decision on which documents to select relies on a neural model evaluating all q, D pairs.

    


In a practical system, it is possible to circumvent some of the issues above by borrowing techniques from the literature on online aggregation literature \cite{online_agg}.
In practice, we can sort the embedding of the images according to the dot product they have with the query. We then process them in batches of predetermined sizes $w$. We stop once a specific tolerance criterion is met, namely when no more than $c$ documents are predicted as relevant by the model.

This leads us to our fourth strategy, which we call Mixed. As the name suggests, we mix two of the strategies already introduced, Neural Selector and TopK. 
Specifically, we take the set union of the TopK (With a small K) and Neural Selector documents to retrieve and to be passed onto the Reasoner.

\subsection{Results}

In this section, we present the experimental evidence to support the ideas presented in this paper.
First, we will show results that test the performance of single architecture components.
Following that, we proceed to evaluate the entirety of our pipeline.
Results for all metrics are reported together with their standard error.

\begin{table*}[]
\caption{Comparison among different retrieval strategies.
The Threshold strategy achieves higher F1 and precision scores, while the ``Mixed" strategy has a higher recall. Once again, we opt for the strategy that achieves higher recall, namely Mixed, as it allows the Reasoner module to receive as much relevant information as possible, ultimately reducing the final pipeline error.}
\label{table:retriever_strat}
\begin{tabular}{@{}lcccccc@{}}
\toprule
\textbf{Selection Strategy} & \multicolumn{1}{c}{\textbf{$\mu$F1}} & \multicolumn{1}{c}{\textbf{$\mu$Recall}} & \multicolumn{1}{c}{\textbf{$\mu$Precision}}  & \multicolumn{1}{c}{\textbf{F1}} & \multicolumn{1}{c}{\textbf{Recall}} & \multicolumn{1}{c}{\textbf{Precision}}  \\ \midrule
Top-K & 0.211 $\pm$ 0.001 & 0.683 $\pm$ 0.004 & 0.125 $\pm$ 0.001 & 0.201 $\pm$ 0.009 & 0.852 $\pm$ 0.018 & 0.125 $\pm$ 0.009 \\
Threshold & $\mathbf{0.351 \pm 0.003}$ & 0.226 $\pm$ 0.003 & $\mathbf{0.791 \pm 0.006}$ & $\mathbf{0.445 \pm 0.029}$ & 0.377 $\pm$ 0.030 & $\mathbf{0.776 \pm 0.022}$ \\
Neural & 0.337 $\pm$ 0.002 & $\mathbf{0.932 \pm 0.002}$ & 0.205 $\pm$ 0.002 & 0.343 $\pm$ 0.016 & 0.898 $\pm$ 0.018 & 0.235 $\pm$ 0.016 \\
Mixed  & 0.337 $\pm$ 0.002 & $\mathbf{0.932 \pm 0.002}$ & 0.205 $\pm$ 0.002 & 0.347 $\pm$ 0.016 & $\mathbf{0.905 \pm 0.015}$ & 0.228 $\pm$ 0.014 \\
\bottomrule
\end{tabular}
\end{table*}

We start by evaluating our retriever strategy. 
We argue that, for our pipeline, a good retriever should have a high level of recall since every relevant document that is failed to be retrieved will produce an error that will propagate to the subsequent components and onto the final response.
For this reason, we explicitly express a preference for models and strategies obtaining a high recall.
We tested each of the 8 CLIP model versions on each of the 4 crafted strategies.
For the sake of space efficiency, we only show results for the various models in the chosen final setting - mixed strategy - and the comparison between different strategies using the best model - ViT-L/14@366px.
In Table \ref{table:retriever}, you can see the performance of the various models in the Mixed Strategy setting.
The first thing we can notice is that while there is a shift in scale between $ \mu $ and macro metrics, at least for precision and recall, the ranking between different models does not really change.
Furthermore, While ViT-L/14@366px is the best model neither with respect to F1 nor precision, it is the best model when considering a recall.
In fact, it consistently beat other models in that regard, with the exception of its twin ViT-L/14, with which the difference in terms of performance is minimal. 
Since the difference in the number of parameters and general complexity is almost unnoticeable, too, we saw no reason not to proceed with the former.
In Table \ref{table:retriever_strat}, we report results for the 4 retrieving strategies we tested.
Once again, while Threshold offers the best precision, the Neural Selector, particularly the Mixed Strategy, offers the best overall results with comparable F1 and much higher Recall.
In Table \ref{table:processor}, we show the difference in performance between the various OFA version we tested.
We only show results for the COUNT query type for the sake of not being repetitive since the difference between these models transfers across tasks.
In this case, unlike the retriever, we see significant differences in results between the model versions tested.
Larger models clearly outperform smaller ones by a wide margin.
Moreover, OFA-huge outperforms OFA-large in terms of total error and $\Delta$ error, while the latter achieves higher accuracy.
We choose OFA-large for two reasons: (i) we favor accuracy over the other two metrics, and (ii) it has half the parameters with respect to the huge version (0.5B vs. 1B).
We also report on a finetuned version of OFA-large (OFA-large FT), obtained by finetuning OFA-large on the train set for 10 epochs with a learning rate of $5e-5$ with the same task. 
Finetuning the OFA model significantly boosts its performance on the MMNDB task.

\begin{table}[h]
\centering
\caption{We test different neural models to be used as the building block for the reasoner on the PerfectIR setting. Smaller models clearly fail to compete with their larger counterparts. OFA-huge achieves a smaller total and $\Delta$ error, while OFA-large has higher accuracy. We choose the latter as we favor accuracy over the other metrics and because it has half the amount of parameters. We also report on a finetuned version that significantly improves over the stock versions.}
\label{table:processor}
\begin{tabular}{lccccccc}
\toprule
\textbf{Model} & \textbf{Total Error $\downarrow$} & \textbf{$\Delta$ Error $\downarrow$} & \textbf{Accuracy $\uparrow$} \\
\hline
OFA-base & 0.831 $\pm$ 0.024 & 2.876 $\pm$ 0.217 & 0.094 $\pm$ 0.014 \\
OFA-medium & 0.871 $\pm$ 0.013 & 2.869 $\pm$ 0.180 & 0.074 $\pm$ 0.005 \\
OFA-large & 0.460 $\pm$ 0.073 & 4.645 $\pm$ 1.944 & $\mathbf{0.597 \pm 0.022}$ \\
OFA-huge & $\mathbf{0.392 \pm 0.025}$ & $\mathbf{2.363 \pm 0.179}$ & $0.533 \pm 0.023$ \\ \hline
{OFA-large FT} & $\mathbf{0.138 \pm 0.011}$ & $\mathbf{1.455 \pm 0.100}$ & $\mathbf{0.668 \pm 0.018}$ \\
\bottomrule
\end{tabular}
\end{table}

The metrics tracked, though spun off, are the same as in the test whose results are reported in Table \ref{table:pipeline_count}.
Here, we test both the reasoner capabilities and the full pipeline.
We perform our testing under 4 different scenarios, considering both the case in which we have a stock model or a finetuned one, reporting on 10 different metrics.
We use the PerfectIR setting as a baseline.
In this setting, the set of documents retrieved $D_r$ is the set of documents that are actually relevant, taken directly from the ground truth. 
This, of course, is an ideal setting in which we assume a perfect retriever and acts as a sort of upper bound for our method.
Full pipeline instead refers to our actual setting, in which our mixed strategy retriever passes the set of retrieved documents.
The metrics we collect are of two kinds: one, with the word total as antecedent, refers to the whole pipeline; the others, without the word total in them, are meant as a test on the intermediate answers $A_p$ produced by the reasoner.
In particular, By accuracy, we mean the percentage of intermediate answers $A_{p_{i}}$that are exactly equal to their ground truth value. 
This is then averaged over all queries.
We then further divide this computation into two disjoint sets, namely, accuracy for true positives (TP), documents in $D_r$ that are actually relevant, and accuracy on false positives (FP), documents in $D_r$ that should not have been retrieved. Please note that in the case of PerfectIR, the set of FP documents is empty by definition.
Since the task at hand is that of the query type COUNT, we are also interested in knowing of close an intermediate answer is to the ground truth value.
We track this with the metric $\Delta$ error. Here, similarly to the accuracy metric, we register the mean absolute deviation between the intermediate answer $A_{p_{i}}$ and the ground truth, averaged over all queries.
Once again, we spun this off into its two components, namely TP and FP.


\begin{table}[h]
\centering
\caption{Results for the query type MAX. It can be immediately noticed how much the finetuning process improves the performance of the MAX query type. In particular, we notice that finetuned models are less prone to produce indecisive intermediate answers such as ``many" and ``a lot", which are highly relevant to this query.
We also notice how close the Full Pipeline setting is to PerfectIR compared to other queries. We argue this is due to the reduced impact of damaging documents, i.e., it is unlikely that a damaging document will be a likely candidate for MAX.}
\label{table:pipeline_max}
\begin{tabular}{lccc}
\toprule

\textbf{Stock} & \textbf{Total Error $\downarrow$} & \textbf{$\Delta$ Error $\downarrow$} & \textbf{Accuracy $\uparrow$}  \\
\midrule
Perfect IR & $\mathbf{2.845 \pm 1.759}$ & $\mathbf{29.263 \pm 17.598}$ & 0.188 $\pm$ 0.044 \\
Noisy IR & 4.576 $\pm$ 2.486 & 41.438 $\pm$ 21.343 & 0.200 $\pm$ 0.045 \\
Dmg. IR & 4.258 $\pm$ 2.035 & 53.325 $\pm$ 23.933 & 0.188 $\pm$ 0.044 \\
Full & 4.280 $\pm$ 2.014 & 53.063 $\pm$ 24.027 & $\mathbf{0.213 \pm 0.046}$ \\ \midrule
\textbf{FTmodel}                   & & &  \\ \midrule
Perfect IR & $\mathbf{0.229 \pm 0.035}$ & 1.813 $\pm$ 0.271 & $\mathbf{0.575 \pm 0.056}$ \\
Noisy IR & $\mathbf{0.229 \pm 0.035}$ & $\mathbf{1.800 \pm 0.273}$ & 0.550 $\pm$ 0.055 \\
Dmg. IR & 0.303 $\pm$ 0.060 & 2.100 $\pm$ 0.320 & 0.525 $\pm$ 0.056 \\
Full & 0.317 $\pm$ 0.056 & 2.263 $\pm$ 0.342 & 0.563 $\pm$ 0.055 \\ \bottomrule
\end{tabular}
\end{table}

Under these two metrics, we can see that the Full Pipeline results are competitive, if not better, with the PerfectIR version.
Upon further inspection, we can also deduct the cause.
In fact, in Full Pipeline, false positive documents are added to the computations. 
Many of these documents are actually easier to deal with since they do not contain the object of interest and can produce an intermediate answer of $0$, raising both the accuracy and the $\Delta$ error of the Full Pipeline version.
In our experimenting, we also noticed that the stock model was struggling to produce useful intermediate results in some instances.
For instance, the model would produce indecisive answers like ``many" and ``few".
Using some prompt engineering, explicitly asking the model to ``Answer with a number" alleviated the problem but did not totally eradicate it.
For this reason, as mentioned earlier, we produced a finetuned version of the reasoner, which improves the accuracy score and dramatically reduces the $\Delta$ error.

Finally, we report results on the total error metric.
Under this metric, we consider the final outcome of the pipeline $o$, and we compute its absolute deviation from the ground truth, averaged over all queries, and normalized by cardinality.
The PerfectIR version achieves excellent results for this task, fully demonstrating the feasibility of the task we propose in this paper.
Full Pipeline, while achieving good scores, lags behind the PerfectIR setting.
To further investigate this difference in performance, we divide the total error into its components.
Once again, TP refers to documents correctly retrieved, FP to documents wrongly retrieved, and false negatives (FN) to documents that should have been retrieved but have not (These last two components are null in the case of PerfectIR by definition).
We notice how the total error TP is actually comparable between the two versions, slightly lower in the case of Full Pipeline since a few of the more challenging documents are not retrieved.
Upon further inspection, we notice that the total error FN is almost negligible, meaning that the gap in total error is not caused by documents not being retrieved.
From the experimental evidence, it is clear that this gap is actually caused by false positives, documents that should not have been retrieved, but they were, nonetheless.

To further investigate this phenomenon, we devise an additional setting called NoisyIR. In this setting, we assume $D_r$ is composed, as in PerfectIR, of the set of relevant documents to which we add, however, some non-relevant documents (300) taken at random.
We notice that the NoisyIR setting performs only slightly worse than the PerfectIR setting, showing that our model is actually robust to noise.

\begin{center}
\begin{table}
\caption{Results for the query type IN. Once again, we observe a gap in performance for the finetuned models. In particular, the finetuned version produces answers that are much more robust to noise.
Moreover, while results are generally satisfactory, we observe an increase in error for the Full Pipeline. We attribute this to damaging documents that trick the reasoner into mispredicting the presence of an object, as evidenced by the high loss for the DamagingIR setting.}
\label{table:pipeline_in}
\begin{tabular}{@{}lcc@{}}
\toprule
{ \textbf{Stock Model}} & \textbf{Total Error $\downarrow$} & \textbf{Accuracy $\uparrow$} \\ \midrule
Perfect IR & $\mathbf{0.131 \pm 0.014}$ & 0.869 $\pm$ 0.014 \\
Noisy IR & 0.404 $\pm$ 0.176 & $\mathbf{0.906 \pm 0.007}$ \\
Damaging IR & 0.829 $\pm$ 0.357 & 0.811 $\pm$ 0.013 \\
Full & 0.793 $\pm$ 0.150 & 0.672 $\pm$ 0.018 \\
\midrule
{\textbf{FTmodel}} &  &  \\ \midrule
Perfect IR & $\mathbf{0.060 \pm 0.007}$ & 0.940 $\pm$ 0.007 \\
Noisy IR & 0.085 $\pm$ 0.007 & $\mathbf{0.946 \pm 0.004}$ \\
Damaging IR & 0.436 $\pm$ 0.054 & 0.838 $\pm$ 0.008 \\ 
Full & 0.330 $\pm$ 0.015 & 0.877 $\pm$ 0.007 \\\bottomrule
\end{tabular}
\end{table}
\end{center}

Following this experiment, we devised a new setting, identical to NoisyIR, but in which the negative documents are not taken at random anymore.
In fact, we take the non-relevant document whose CLIP embedding with the query is the highest.
We call this setting DamagingIR.
Results clearly show that these documents are able to "trick" the reasoner into generating wrong intermediate answers, causing a large FP error and ultimately a more significant total error resulting in a performance difference between the PerfectIR version and the Full Pipeline one.

DamagingIR has already been observed by \cite{sauchuk2022role} and, to the best of the authors' knowledge, has not been yet fully addressed. At the end of this Section, we provide a more complete commentary on this issue.

In Table \ref{table:pipeline_in}, we show results for the IN query type.
This query answers questions of the type ``In how many pictures there are \{object\}?".
We consider two metrics in this scenario that mirror the ones defined for the COUNT setting.
First, we consider accuracy, that is, the percentage of time the intermediate results $A_{p_{i}}$ are exactly equal to their respective ground truths.
The total error indicates the absolute deviation of the total number of documents found satisfying the condition from its ground truth, later averaged over all queries and normalized by cardinality.
We can immediately notice that the finetuned version of the reasoner generally performs better with respect to its stock counterpart.
We also notice the positive results obtained by the Full Pipeline, even though they are lower than the near-perfect PerfectIR.
Once again, even more clearly than before, we can attribute this reduction in performance to DamagingIR, that is, to false positive documents that manage to ``trick" the model into thinking that there is an object in the image when there is really not, as evidence by the drop in performance observed under this regime.

Finally, we report results for the MAX query type, which return the document with the max instances of a particular object in the collection.
We test on the same 4 scenarios and report on three metrics.
$\Delta$ and total error are specular to previous settings, while total accuracy is the percentage of queries in which the correct document is found.
This is the scenario that shows the most significant difference between the stock reasoner and its finetuned version.
We attribute this gap to an issue cited earlier, in which for pictures with high instances of a particular object, the model would produce indecisive answers like ``many", a problem that the finetuned model does not feature.
Furthermore, we notice that the difference between the PerfectIR and the Full Pipeline version is rather small.
This stems from the fact that, unlike in the two other scenarios, false positives documents are unlikely to be appetible candidates for the MAX type of query, failing to impact the final outcome.
We also register that, even when the model is not able to retrieve the correct max document, the picture found has a comparable number of instances, as indicated by the total error.

Overall, the results are very promising and fully show the potential for Multimodal Neural Databases.
We managed to build an effective and efficient retrieval system with a high recall.
The reasoner module, and the pipeline as a whole, show good performance and resistance to noise, with low error and high accuracy, coupled with a resistance to noise.
However, like other systems in IR, it is weak to DamagingIR, as shown by the increased caused false positives.
We argue that by tackling this issue we can further increase the performance of MMNDB and bring it close to the optimum.

\section{Future Research Directions}
\label{sec:futureR}
The introduction of Multimodal Neural Databases paves the way toward new and exciting research directions; in this section, we proceed to discuss some of the more interesting ones.

In this paper, we have shown the feasibility of the proposed task but have yet to explore many open problems.

First and foremost, a key feature in any database system is the ability to update its information.
In a typical database system, one would expect to be able to remove, add, or modify the information as he wishes.
This is not straightforward under our current paradigm and needs more research efforts.

On this line, it would be crucial to account for the importance of time in databases. 
I could ask the database question like "What is the place I visited the most between 1 pm and 3 pm this year?"

Furthermore, we have restricted ourselves to only two modalities, and in particular, a database made of strictly images.
Expanding available modalities is a clear path with obvious benefits.
Additionally, we could consider not only documents but documents and their meta-data.
To provide an example, whenever we take a picture with our smartphone, we collect a variety of information, such as the location and time, which would definitely be helpful for a database of this kind.

To remain in the field of smartphones, recently, video-clip sharing has become very popular among social network users.
Asking database-like queries on videos is an open problem that presents many challenges.
Among all, it is crucial to be able to identify entities along frames to be able to answer queries effectively.
While recognizing an entity (like a person) is generally feasible for text, it is much more complex when considering different modalities. 
Solving this will be critical for the development of MMNDBs.

In our presentation, we stressed the fact that the proposed architecture is not the only possible way of solving this problem.
In fact, recently, we have witnessed the power of large foundational models to solve a wide array of tasks, with chatGPT and GPT-x models, in general, leading the way \cite{radford2019language}.
We believe that these large foundational models could bring an advance to this field as well.
However, this is not straightforward, and some issues should be addressed.
These models require a large amount of data to be pre-trained; this begs the question of how one could encapsulate the memory used during training from the actual Multimodal Database to avoid knowledge contamination.
By knowledge contamination, we mean the known phenomenon for which data used during pretraining is spilled when generating answers in a completely unrelated context.
 Knowledge contamination proved troublesome in many applications, with some systems allegedly revealing private keys or even personal phone numbers.
Furthermore, true multimodality in these large models remains an open research direction and a major roadblock toward conversational multimodal systems.

Finally, we have taken Multimodal Neural Database in its most general setting.
However, one might be interested in specific scenarios with more precise guidelines and goals.
For instance, there may be cases in which one has a precise idea of which kind of queries are to be expected.
In that case, strategies could be crafted to optimize the system.
In traditional database systems, for example, indexing or creating views for common queries is a prevalent practice.
Creating equivalent procedures for MMNDB is still unexplored.

\section{Related Work}
\textbf{Multimedia Information Retrieval (MMIR)} Bridging the gap between multimodal unstructured data and structured database systems has always been a central key endeavor in Information Retrieval \cite{halevy2003crossing}. The former is vastly highly available on the web but challenging to digest and query compared to the latter. 
Particular focus has been posed on content-based image retrieval \cite{895972, lew2006content, RUI199939} and recently on cross-modal retrieval \cite{wang2016comprehensive, hu2019scalable}, which have been made possible with the recent advancements in deep learning \cite{lecun2015deep}. Specifically, there has been an explosion of such approaches for Image-text retrieval \cite{rao2022does, cheng2022vista, yu2022u, yu2022cross, yu2021heterogeneous, qu2021dynamic}. 
However, these systems are primarily concerned with retrieving relevant documents (e.g., images) based on a given query (e.g., text). In contrast, MMNDBs focus on answering database-like queries on large data collections, which current cross-modal retrieval methods cannot achieve.

\noindent\textbf{Multimodal Neural Models} There has been a recent surge in the development of multimodal neural models that can handle data in different forms, primarily images, and text, for various applications. Usually, this is performed via a single neural multimodal encoder \cite{wang2022unifying, li2023blip, bao2021vlmo, wang2022image, chen2022pali} or via different encoders for each modality that is jointly aligned via a shared space \cite{yu2022coca, radford2021learning}. 
In MMNDBs, we take advantage of this characteristic by using a separate encoder system as a Retriever to precompute and index visual tokens, thus reducing computation and time at runtime by only using the text encoder to compute the textual embedding of the query. However, directly applying these neural models to the MMNDB task would not be scalable due to the high computational cost. We use them as components in our architecture, building on their successes in other vision-language tasks.

\noindent\textbf{Visual Question Answering (VQA)}
Most of these multimodal vision-text models are evaluated on the task of visual question answering \cite{goyal2017making}, where the goal is to generate an accurate and semantically coherent response based on a question about an image. Usually, these involve using reasoning and other capacities that are non-trivial, even for current neural architectures. 
Compared to the task of MMNDBs, VQA is defined on a single image-question pair and requires reasoning over the image to answer the question.
Closer to the task of MMNDBs, is Open-domain Question Answering (OpenQA) \cite{ DBLP:journals/corr/abs-2101-00774} and the multimodal variant WebQA \cite{DBLP:journals/corr/abs-2109-00590} which aim to answer natural language questions over large-scale unstructured textual documents. Compared to the task of MMNDBs, their scope is different and involves multimodal, open-domain question-answering, while we want to focus on efficiently answering database-like queries over a collection of documents in different formats (e.g., images).








\noindent\textbf{Answering Database Queries}
There has been substantial effort put into converting queries expressed in natural language into SQL queries for databases with known structure \cite{androutsopoulos1995natural, li2014constructing, zeng2020photon}, and there have also been advancements in adapting this approach for semi-structured data and knowledge bases \cite{berant2013semantic, pasupat2015compositional}.

Recently, \citet{thorne2021database, thorne2021natural} proposed NeuralDB as a way to perform database queries over a collection of textual documents without the need to translate data or queries into a predefined database schema but using parallel neural techniques instead. Their approach is very effective but it: (i) requires preprocessing and analysis for the aggregation operator; (ii) is limited to simple queries and (iii) is capable of handling data just in textual format. In this paper, we stem from this research approach and tackle the third limitation extending the original architecture proposed to multimodal document processing.



\noindent\textbf{Retrieval-augmented models}
Recently there has been a surge of interest in the line of research concerning retrieval-augmented neural models \cite{10.1145/3477495.3532682}. Most of the current models focus on augmenting current language models' capabilities with an external memory or retrieval mechanism that retrieves relevant documents given an input query, reducing the number of parameters and non-factual errors \cite{https://doi.org/10.48550/arxiv.2302.07842}. 




\section{Conclusion}

In this paper, we have proposed to expand the field of Multimedia Information retrieval through the introduction of Multimodal Neural Databases.
MMNDBs promise to answer complex database-like queries that involve reasoning over multiple modalities at scale.
We have demonstrated the feasibility and potential of this system by proposing a first principled approach to solve this problem with an architecture composed of three modules - retriever, reasoner, and aggregator - and performing a rich set of experiments.
We have discussed potential future research directions that could stem from the system introduced in this paper.
MMNDBs set a new research agenda that strives to simultaneously act as a bridge between information retrieval and database systems and reduce the gap between the two.
We believe MMNDBs have the potential to substantially advance not only the field of MMIR but the general field of Information Retrieval in its entirety.

\section{Acknowledgment}
This work was partially supported by projects FAIR (PE0000013) and SERICS (PE00000014) under the MUR National Recovery and Resilience Plan funded by the European Union - NextGenerationEU and by ERC Starting Grant No. 802554 (SPECGEO) and PRIN 2020 project n.2020TA3K9N "LEGO.AI".

\clearpage
\newpage
\bibliographystyle{ACM-Reference-Format}
\balance
\bibliography{biblio}




\end{document}